\documentclass[conference,letterpaper]{IEEEtran}
\IEEEoverridecommandlockouts

\usepackage{amsthm,amssymb}
\usepackage{graphicx}
\usepackage{braket}
\usepackage[dvipsnames]{xcolor}
\usepackage{arydshln}
\usepackage[utf8]{inputenc} 
\usepackage[T1]{fontenc}
\usepackage{ifthen}
\usepackage{cite}
\usepackage[cmex10]{amsmath} 
\usepackage{bm}
\usepackage{amsfonts}
\usepackage{xcolor}
\usepackage{bbm}
\usepackage{pgfplots}
\usepackage{colortbl}
\usepackage{pgfplots}
\usepackage{tikz}
\usepackage{tikzpeople}
\usepackage{enumitem}

\newtheorem{theorem}{Theorem}

\newtheorem{remark}{Remark}
\newtheorem{lemma}{Lemma}
\newtheorem{example}{Example}

\newenvironment{Proof}[1]{\medskip\par\noindent{\bf Proof.\,}\,#1}{{\mbox{\,$\blacksquare$}\par}}

\newcommand{\diag}{{\rm{diag}}}

\newcommand{\defeq}{\stackrel{\Delta}{=}}

\pgfplotsset{compat=1.18}
\usetikzlibrary{tikzmark, shapes, fit,backgrounds}
\usetikzlibrary{matrix, decorations.pathreplacing, angles, quotes}
\usetikzlibrary{patterns, calc, positioning}
\usetikzlibrary{arrows, arrows.meta}
\usetikzlibrary{shapes.multipart}
\usetikzlibrary{fit}
\usetikzlibrary{shapes.geometric, positioning}
\usetikzlibrary{decorations.pathmorphing}
\tikzstyle{block} = [rectangle, draw, minimum width=4em, text centered, rounded corners, minimum height=1em]
\tikzstyle{line} = [draw, -latex]
\tikzset{meter/.append style={draw, inner sep=10, rectangle, font=\vphantom{A}, minimum width=30, scale=.7, path picture={\draw[black] ([shift={(.1,.3)}]path picture bounding box.south west) to[bend left=50] ([shift={(-.1,.3)}]path picture bounding box.south east);\draw[black,-{Latex[scale=.5]}] ([shift={(0,.1)}]path picture bounding box.south) -- ([shift={(.3,-.1)}]path picture bounding box.north);}}}
\tikzset{snake it/.style={decorate, decoration=snake}}
\usetikzlibrary{shapes.multipart}
\usetikzlibrary{fit}
\usetikzlibrary{shapes.geometric,positioning}
\usetikzlibrary{calc}
\usetikzlibrary{patterns}
\usetikzlibrary{matrix}
\usetikzlibrary{tikzmark}

\begin{document}

\title{Simultaneously Minimizing Storage and Bandwidth Under Exact Repair With Quantum Entanglement}

\author{Lei Hu \quad Mohamed Nomeir \quad Alptug Aytekin \quad Sennur Ulukus\\
\normalsize Department of Electrical and Computer Engineering\\
\normalsize University of Maryland, College Park, MD 20742 \\
\normalsize \emph{leihu@umd.edu} \quad \emph{mnomeir@umd.edu} \quad \emph{aaytekin@umd.edu}  \quad \emph{ulukus@umd.edu}}

\maketitle

\vspace{-0.5mm}
\enlargethispage{-0.02in}
\begin{abstract}
    We study exact-regenerating codes for entanglement-assisted distributed storage systems. Consider an $(n,k,d,\alpha,\beta_{\mathsf{q}},B)$ distributed system that stores a file of $B$ classical symbols across $n$ nodes with each node storing $\alpha$ symbols. A data collector can recover the file by accessing any $k$ nodes. When a node fails, any $d$ surviving nodes share an entangled state, and each of them transmits a quantum system of $\beta_{\mathsf{q}}$ qudits to a newcomer. The newcomer then performs a measurement on the received quantum systems to generate its storage. Recent work \cite{Hu_breaking} showed that, under functional repair where the regenerated content may differ from that of the failed node, there exists a unique optimal regenerating point that \emph{simultaneously minimizes both storage $\alpha$ and repair bandwidth $d \beta_{\mathsf{q}}$} when $d \geq 2k-2$. In this paper, we show that, under \emph{exact repair}, where the newcomer reproduces exactly the same content as the failed node, this optimal point remains achievable. Our construction builds on the classical product-matrix framework and the Calderbank-Shor-Steane (CSS)-based stabilizer formalism.
\end{abstract}

\section{Introduction}
Distributed storage systems store a file across multiple nodes with redundancy to ensure reliability under node failures \cite{Rhea_storage, Bhagwan_storage}. Specifically, a file consisting of $B$ symbols (dits) is encoded and distributed over $n$ storage nodes, each storing $\alpha$ dits, such that a user or data collector (DC) can reconstruct the file by accessing any $k$ nodes (the data-retrieval property). To minimize storage overhead, one can employ maximum distance separable (MDS) codes, where each node stores $\alpha = B/k$ dits and any $k$ nodes suffice to recover the entire file. However, this optimal storage efficiency comes at the cost of high repair bandwidth: regenerating a failed node requires downloading $B$ dits from $k$ nodes, each node sending $\beta = B/k$ dits, and then the communication cost is $k$ times the per-node storage \cite{Suh_exact}. This leads to significant inefficiency in terms of communication cost. Indeed, by allowing the newcomer to contact $d \geq k$ helper nodes, the download cost can be reduced to $\frac{B}{k}\cdot \frac{d}{d-k+1}$, which gives a reduction factor of $\frac{k(d-k+1)}{d}$. For example, when $(n,k,d) = (21,10,20)$, we can have a more than $5$x reduction in communication cost. This reveals a fundamental tradeoff between storage $\alpha$ and repair bandwidth $d \beta$: decreasing storage necessarily increases repair bandwidth, and vice versa.

Recently, quantum entanglement has been shown to be able to improve the transmission efficiency of classical information in distributed systems \cite{entanglement_assisted}, especially for private information retrieval \cite{aytekin2023quantum,nomeir2025byzantine,jafar_quantum_unresponsive} and distributed computation \cite{yao2024inverted,entanglement_assisted}, where the communication cost can be reduced by half compared to the classical setting. In distributed storage systems, entanglement-assisted repair also helps improve the tradeoff between storage and repair bandwidth \cite{Hu_breaking}. Specifically, when a node fails, the helper nodes share prior entanglement, encode classical information into quantum systems, and transmit the quantum systems to the newcomer. The newcomer then performs measurements to recover classical information and regenerate its storage, so that the data-retrieval property continues to hold. This framework can significantly improve the classical storage-bandwidth tradeoff \cite{Hu_breaking,QRC_arxiv}. Surprisingly, when $d \geq 2k-2$, there exists a unique optimal regenerating point that simultaneously minimizes both storage and repair bandwidth. This phenomenon does not appear in the classical setting. 

However, these results \cite{Dimakis_2010,Hu_breaking,QRC_arxiv} are established under functional repair, where the regenerated content may differ from the failed node’s original content as long as the data-retrieval property is satisfied. From a practical perspective, functional repair introduces additional system overhead: encoding coefficients must be updated and communicated across the network, and both data retrieval  and future repairs must adapt accordingly \cite{Exact_regeneration,Suh_exact}. This increases communication and computational complexity. Consequently, exact repair, where the newcomer reconstructs the exact content of the failed node, is often preferred in practice. However, exact repair imposes stricter constraints for the system, and it is unclear in general whether the optimal tradeoff achieved under functional repair can also be attained under exact repair. Even in the classical setting, this question remains unresolved in general \cite{tian2014characterizing}.

As a first step toward characterizing the fundamental limit of storage-bandwidth in entanglement-assisted distributed storage systems under exact repair, in this paper, we show that the optimal regenerating point for $d \geq 2k-2$ is \emph{achievable}. Our construction builds on the classical product–matrix framework, combined with the Calderbank-Shor-Steane (CSS)-based stabilizer implementation \cite{calderbank1996good,steane1996multiple}, where a transfer matrix satisfying the dual containing condition is carefully designed to enable the repair process.

\enlargethispage{-0.01in}
\textit{Notation:}
For a positive integer $M$, $[M]$ denotes the set $\{1, \ldots, M\}$. For integers $a$ and $b$, $[a:b]$ denotes $\{a,\ldots,b \}$. $A_{[M]}$ is the compact notation of $\{A_{1}, A_2, \ldots, A_{M}\}$. $\mathbb{F}_{q}$ denotes a finite field with order $q$, where $q = p^r$ with $p$ a prime number and $r$ a positive integer. The symbol $\defeq$ means ``defined as''. $\mathbb{F}_{q}^{a \times b}$ is the set of $a \times b$ matrices with elements in $\mathbb{F}_{q}$. $\mathbf{I}_{a}$ is an $a \times a$ identity matrix, and $\mathbf{0}_{a \times b}$ is an $a \times b$ all-zero matrix. $\mathrm{blkdiag}(\cdot)$ is a block matrix where all off-diagonal blocks are zero matrices. For a matrix $\mathbf{A}$, $\mathrm{rowspan}(\mathbf{A})$ denotes the vector subspace spanned by the rows $\mathbf{A}$. $\mathbf{A}^T$ denotes the transpose of $\mathbf{A}$. $[\mathbf{A};\mathbf{B}]$ denotes the vertical concatenation of matrices $\mathbf{A}$ and $\mathbf{B}$.
For a linear code $\mathcal{C} \subseteq \mathbb{F}_q^n$, its dual is defined as
\begin{align}
    \mathcal{C}^{\perp} \triangleq \left\{ \mathbf{v} \in \mathbb{F}_q^n : \mathbf{v}^T \mathbf{c} = 0,\ \forall\, \mathbf{c} \in \mathcal{C} \right\}.
\end{align}
Vectors are column vectors unless otherwise specified.

\begin{figure*}[t]
    \centering
    \includegraphics[width=0.8\textwidth]{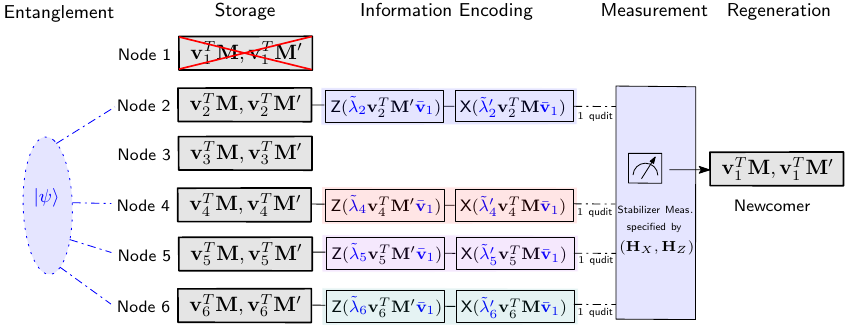}
    \caption{The entanglement-assisted distributed storage system under exact repair when $(n,k,d) = (6,3,4)$, $\alpha=2$, and $\beta_{\mathsf{q}} = 1$ in Example~\ref{example1}.}
    \label{fig:thm1}
\end{figure*}

\enlargethispage{-0.01in}
\section{Problem Statement}
\subsection{Distributed Storage System with Exact Repair}
Consider an $(n,k,d,B,\alpha,\beta_{\mathsf{q}})$ distributed storage system with entanglement-assisted repair. A file with $B$ dits over $\mathbb{F}_q$ is distributed across $n$ storage nodes, each storing $\alpha$ dits. A DC can retrieve the file from any subset of $k$ nodes; this is referred to as the \textit{data-retrieval property}.

In addition, we focus on the exact repair setting. When a node $f \in [n]$ fails, a newcomer connects to any $d$ surviving nodes and aims to reconstruct \emph{exactly} the same content as that stored in node $f$. In the repair process, we consider quantum communication assistance. Specifically, the $d$ helper nodes are allowed to share prior entanglement. Each helper node transmits $\beta_{\mathsf{q}}$ qudits to the newcomer, who performs a quantum measurement on the received systems to generate $\alpha$ classical symbols for storage. The entanglement-assisted repair process consists of the following three stages:
\begin{enumerate}
    \item \textbf{(Entanglement preparation)}
    Suppose node $f$ fails, and the newcomer connects to a set of helper nodes $s_{[d]} \defeq \{s_1,\ldots,s_d\} \subseteq [n]\setminus {f}$. The helpers initially share an entangled quantum system $\mathcal{Q} = \mathcal{Q}_1  \cdots \mathcal{Q}_d$ over a $q$-dimensional system. The state of the system is described by a density operator $\rho^{\mathrm{ini}}$. Each subsystem $\mathcal{Q}_s, s \in s_{[d]}$ has dimension $\delta_s$, corresponding to $\log_q \delta_s = \beta_{\mathsf{q}}$ qudits.
    
    \item \textbf{(Encoding at helper nodes)}
    Each helper node $s \in s_{[d]}$ applies a local encoding operation on its subsystem $\mathcal{Q}_s$, which only depends on its stored data. Specifically, node $s$ applies a completely positive trace preserving mapping (CPTP) map $\Lambda_s$ to its subsystem. After all encoding operations are applied, the joint state becomes
    \begin{align}
        \rho^{\rm enc} = \left( \bigotimes_{s \in s_{[d]} } \Lambda_s \right)  \left(\rho^{\rm ini} \right).
    \end{align}
    The resulting quantum subsystems are then transmitted to the newcomer through noiseless quantum channels. The repair bandwidth (total communication cost) is then $d \beta_{\mathsf{q}}$ qubits.
    
    \item \textbf{(Exact generation)}
    Upon receiving the quantum systems, the newcomer performs a quantum measurement (POVM) on the joint state. Based on the measurement outcome, the newcomer generates $\alpha$ dits, which exactly reproduce the content originally stored in the failed node $f$.
\end{enumerate}

\subsection{Existing Results}
In classical distributed storage systems, during the repair process each helper node transmits $\beta_{\mathsf{c}}$ dits to the newcomer. The fundamental storage-repair bandwidth tradeoff is characterized by \cite{Dimakis_2010}
\begin{align}
    \mathrm{[Classical]}: \sum_{i=0}^{k-1} \min \{(d-i)\beta_{\mathsf{c}}, \alpha\} \geq B.
\end{align}

In the case of entanglement-assisted repair, \cite{Hu_breaking,QRC_arxiv} show that the storage-bandwidth tradeoff is improved to
\begin{align}
    \mathrm{[Quantum]}: \sum_{i=0}^{k-1} \min\{2(d-i)\beta_{\mathsf{q}}, d\beta_{\mathsf{q}}, \alpha\} \geq B. \label{eq:quantum_tradeoff}
\end{align}
A notable consequence is that when $d \geq 2k-2$, the tradeoff admits a \emph{single optimal regenerating point} that simultaneously minimizes both storage and repair bandwidth, given by
\begin{align}
    (\alpha^{*}, d\beta_{\mathsf{q}}^{*}) = \left(\frac{B}{k}, \frac{B}{k}\right).\label{eq:optimal}
\end{align}
However, this result holds under \emph{functional repair}, where the regenerated content is not required to be identical to that of the failed node.

In this paper, we will show that such an optimal point is also achievable in the \emph{exact regeneration} setting. To this end, we first present a framework for transmitting classical information over quantum systems, based on the CSS stabilizer formalism \cite{calderbank1996good,steane1996multiple}.

\enlargethispage{-0.01in}
\subsection{CSS-based Stabilizer Formalism \cite{calderbank1996good,steane1996multiple}}
A CSS code $\mathrm{CSS}(\mathcal{C}_X, \mathcal{C}_Z)$ is specified by two classical linear codes $\mathcal{C}_X,\mathcal{C}_Z\subseteq \mathbb{F}_q^N$ satisfying the dual containment condition $\mathcal{C}_X^\perp \subseteq \mathcal{C}_Z$ \cite{gottesman2024surviving}. Let $\mathbf{H}_X \in \mathbb{F}_q^{r_X \times N}$ and $\mathbf{H}_Z \in \mathbb{F}_q^{r_Z \times N}$ denote the parity-check matrices of $\mathcal{C}_X$ and $\mathcal{C}_Z$, respectively. Let $\mathbf{x} \defeq [x_1,\ldots,x_N]^T$, $\mathbf{z} \defeq [z_1,\ldots,z_N]^T \in \mathbb{F}_q^N$.
For a Pauli error of the form
\begin{align}
    \mathsf{X}(\mathbf{x})\mathsf{Z}(\mathbf{z}) \defeq \mathsf{X}(x_1)\mathsf{Z}(z_1) \otimes \cdots \otimes \mathsf{X}(x_N)\mathsf{Z}(z_N),
\end{align}
the corresponding syndromes obtained from stabilizer measurements are given by \cite{lu2025quantum}
\begin{align}
    \mathbf{s}_X = \mathbf{H}_Z \mathbf{x}, \quad \mathbf{s}_Z = \mathbf{H}_X \mathbf{z}.\label{eq:syndrome}
\end{align}

\begin{remark}
    The vectors $\mathbf{s}_X$ and $\mathbf{s}_Z$ correspond to the syndromes associated with $X$-type and $Z$-type errors, respectively. Notably, the syndromes are linear functions of the classical data vectors $\mathbf{x}$ and $\mathbf{z}$. This structure enables the transmitters to encode their classical data through $\mathsf{X}$ and $\mathsf{Z}$ type Pauli operations, such that the resulting syndromes correspond to linear combinations of the data. 
    
    In particular, consider a distributed setting with $N$ transmitters sharing an entangled state $\ket{\psi} \in \mathrm{CSS}(\mathcal{C}_X, \mathcal{C}_Z)$. Each transmitter $i \in [N]$ encodes its local classical data $(x_i,z_i) \in \mathbb{F}_q^{2}$ by applying the Pauli operator $\mathsf{X}(x_i)\mathsf{Z}(z_i)$ to its local state. Upon receiving the joint quantum state, the receiver performs stabilizer measurements and obtains the syndromes in \eqref{eq:syndrome}. By appropriately designing the parity-check matrices $\mathbf{H}_Z$ and $\mathbf{H}_X$, the receiver can recover desired linear combinations of the distributed data across the transmitters. This observation enables the achievability construction developed in this work.
\end{remark}

\enlargethispage{-0.02in}
\section{Main Result and an Example}
\begin{theorem}\label{thm_main}
    Consider an $(n,k,d,\alpha,\beta_{\mathsf{q}},B)$ distributed storage system under exact repair with entanglement-assisted quantum communication. If $d \geq 2k-2$, then the regenerating point
    \begin{align}
        \left( \alpha = \frac{B}{k}, \  d\beta_{\mathsf{q}} = \frac{B}{k} \right)
    \end{align}
    is achievable and optimal. In particular, this point simultaneously minimizes storage and repair bandwidth.
\end{theorem}

\begin{remark}
    The optimality follows from the converse bound established for functional repair, where the point $\alpha=d\beta_{\mathsf{q}}=B/k$ is already known to be optimal. Since exact regeneration imposes a strictly stronger requirement than functional repair, any point strictly improving upon it is not achievable.
    
    To establish achievability, we develop a coding framework based on classical product-matrix constructions and CSS-based stabilizer formalism. Specifically, we adopt the product-matrix code to design the storage such that the data-retrieval property is satisfied. During repair, the helper nodes encode their stored classical symbols by applying Pauli operators on a shared entangled state, following the CSS stabilizer formalism \cite{calderbank1996good,steane1996multiple}. The resulting stabilizer measurements at the newcomer yield carefully designed linear combinations of the stored data, which are sufficient to exactly reconstruct the content of the failed node. 
    
    To build intuition for the construction, we next illustrate the key ideas through a concrete example.
\end{remark}

\begin{example}\label{example1}
    As shown in Fig. \ref{fig:thm1}, consider $(n,k,d) = (6,3,4)$ with $\beta_{\mathsf{q}} = 1$. Let $B = 12$, then at the optimal regenerating point, $\alpha = d\beta_{\mathsf{q}} = B/k =4$.
    
    \textbf{Storage design:}
    We work over the finite field $\mathbb{F}_{13}$. The $12$ data symbols are denoted by two instances $u_{[6]}$ and $u_{[6]}'$, and are rearranged into symmetric matrices
    \begin{align}
        &\mathbf{S}_1 \defeq \begin{bmatrix}
            u_1 & u_2\\
            u_2 & u_3
        \end{bmatrix},\quad 
        \mathbf{S}_2 \defeq \begin{bmatrix}
            u_4 & u_5\\
            u_5 & u_6
        \end{bmatrix},\\
        & \mathbf{S}_1' \defeq \begin{bmatrix}
            u_1' & u_2'\\
            u_2' & u_3'
        \end{bmatrix},\quad 
        \mathbf{S}_2' \defeq \begin{bmatrix}
            u_4' & u_5'\\
            u_5' & u_6'
        \end{bmatrix}.
    \end{align}
    Define the stacked matrices $\mathbf{M}= [\mathbf{S}_1;\mathbf{S}_2]$, $\mathbf{M}' = [\mathbf{S}_1';\mathbf{S}_2'] \in \mathbb{F}_{13}^{4 \times 2}$. The storage matrix is constructed as
    \begin{align}
        \mathbf{C} = \begin{bmatrix}
            \mathbf{V} & \mathbf{0}_{6 \times 4} \\
            \mathbf{0}_{6 \times 4} & \mathbf{V}
        \end{bmatrix}
        \begin{bmatrix}
            \mathbf{M} \\
            \mathbf{M}'
        \end{bmatrix}
        =
        \begin{bmatrix}
            \mathbf{V} \mathbf{M}\\
            \mathbf{V} \mathbf{M}'
        \end{bmatrix} \in \mathbb{F}_{13}^{12 \times 2},
    \end{align}
    where $\mathbf{V}\in \mathbb{F}_{13}^{6 \times 4}$ is a Vandermonde matrix defined as
    \begin{align}
        \mathbf{V} \defeq 
        \begin{bmatrix}
            1 & 1 & 1 & 1\\
            1 & 2 & 4 & 8\\
            1 & 3 & 9 & 1 \\
            1 & 4 & 3 & 12 \\
            1 & 5 & 12 & 8 \\
            1 & 6 & 10 & 8 
        \end{bmatrix}.
    \end{align}
    Equivalently,
    \begin{align}
        \mathbf{V} = 
        \begin{bmatrix}
            \bar{\mathbf{V}} & \mathbf{\Lambda} \bar{\mathbf{V}}
        \end{bmatrix},
    \end{align}
    where $\bar{\mathbf{V}}$ is the matrix consisting of the first two columns of $\mathbf{V}$, and $\mathbf{\Lambda} \defeq \diag(\boldsymbol{\lambda})$,
    \begin{align}
        \boldsymbol{\lambda} = 
        \begin{bmatrix}
            1& 4& 9& 3& 12 & 10
        \end{bmatrix}^T.
    \end{align}

    Let $\mathbf{v}^T_i$ denote the $i$th row of $\mathbf{V}$. Node $i$ stores the $i$th and $(i+n)$th rows of $\mathbf{C}$, denoted by
    \begin{align}
        \bar{\mathbf{c}}_i^T \defeq 
        \begin{bmatrix}
            \mathbf{C}(i,:) &  
            \mathbf{C}(i+n,:)
        \end{bmatrix}
        = \begin{bmatrix}
            \mathbf{v}_i^T \mathbf{M} & 
            \mathbf{v}_i^T \mathbf{M}'
        \end{bmatrix}.
    \end{align}
    For instance, node $1$ stores
    \begin{align}
        \bar{\mathbf{c}}_1 = \begin{bmatrix}
            u_1 + u_2 + u_4 + u_5 \\
            u_2 + u_3 + u_5 + u_6 \\
            u_1' + u_2' + u_4' + u_5'\\
            u_2' + u_3' + u_5' + u_6'
        \end{bmatrix}.\label{eq:c_1}
    \end{align}
    The data-retrieval property follows from the product-matrix construction in \cite{Exact_regeneration}.

    \textbf{Single-node repair:}
    Assume that node $1$ fails and the helper nodes are $\{2,4,5,6\}$. Each node $i \in \{2,4,5,6\}$ computes 
    \begin{align}
        \tilde{\lambda}_i \mathbf{v}_i^T \mathbf{M} \bar{\mathbf{v}}_1, \quad \tilde{\lambda}_i'
        \mathbf{v}_i^T \mathbf{M}' \bar{\mathbf{v}}_1,
    \end{align}
    where 
    \begin{align}
        (\tilde{\lambda}_2,\tilde{\lambda}_4,\tilde{\lambda}_5,\tilde{\lambda}_6) &= (9,7,6,3), \\
        (\tilde{\lambda}_2',\tilde{\lambda}_4',\tilde{\lambda}_5',\tilde{\lambda}_6') &=(11,5,11,2),
    \end{align}
    are precoding scalars. Collect the corresponding $8$ classical symbols into
    \begin{align}
        \mathbf{y} = \begin{bmatrix}
            \tilde{\mathbf{\Lambda}} & \mathbf{0} \\
            \mathbf{0} & \tilde{\mathbf{\Lambda}}'
        \end{bmatrix}
        \underbrace
        {\begin{bmatrix}
            \tilde{\mathbf{V}} & \mathbf{0} \\
            \mathbf{0} & \tilde{\mathbf{V}}
        \end{bmatrix}
        \begin{bmatrix}
            \mathbf{M}\\
            \mathbf{M}'
        \end{bmatrix}}_{\mathrm{storage}}
        \bar{\mathbf{v}}_1 \in \mathbb{F}_{13}^{8 \times 1},\label{eq:ex1_y}
    \end{align}
    where $\tilde{\mathbf{V}}$ consists of the rows in $\mathbf{V}$ corresponding to the helper nodes, i.e.,
    \begin{align}
        \tilde{\mathbf{V}} \defeq [\mathbf{v}^T_2;\mathbf{v}^T_4;\mathbf{v}^T_5;\mathbf{v}^T_6] = 
        \begin{bmatrix}
            1 & 2 & 4 & 8\\
            1 & 4 & 3 & 12\\
            1 & 5 & 12 & 8\\
            1 & 6 & 10 & 8
        \end{bmatrix},
    \end{align}
    $\tilde{\mathbf{\Lambda}} \defeq \diag(\tilde{\lambda}_2,\tilde{\lambda}_4,\tilde{\lambda}_5,\tilde{\lambda}_6)$, and $\tilde{\mathbf{\Lambda}}' \defeq \diag(\tilde{\lambda}_2',\tilde{\lambda}_4',\tilde{\lambda}_5',\tilde{\lambda}_6')$.

    Before transmission, the helper nodes $2,4,5,6$ share an entangled state $\ket{\psi}$ in the codespace $\mathrm{CSS}(\mathcal{C}_X, \mathcal{C}_Z)$ \cite{calderbank1996good,steane1996multiple}, where the associated parity-check matrices are 
    \begin{align}
        \mathbf{H}_X & = \begin{bmatrix}
            \mathbf{I}_2 & \mathbf{I}_2
        \end{bmatrix}
        (\tilde{\mathbf{\Lambda}}' \tilde{\mathbf{V}})^{-1} = \begin{bmatrix}
            11 & 10 & 3 & 9\\
            4 & 2 & 1 & 0
        \end{bmatrix},\\
        \mathbf{H}_Z & = \begin{bmatrix}
            \mathbf{I}_2 & \mathbf{I}_2
        \end{bmatrix}
        (\tilde{\mathbf{\Lambda}} \tilde{\mathbf{V}})^{-1} = \begin{bmatrix}
            12 & 9 & 12 & 6 \\
            2 & 7 & 4 & 0
        \end{bmatrix}.
    \end{align}
    
    Each helper node encodes its computed symbols in \eqref{eq:ex1_y} into its local quantum subsystem by applying Pauli operators. Specifically, the helper node $i$ applies $\mathsf{X}(y_i)\mathsf{Z}(y_{i+4})$ on its local quantum state, and the operation on the whole entangled state is then represented by
    \begin{align}
        \rho^{\rm enc} = \mathsf{X}(\mathbf{y}_1)\mathsf{Z}(\mathbf{y}_2) \ket{\psi}\bra{\psi} (\mathsf{X}(\mathbf{y}_1)\mathsf{Z}(\mathbf{y}_2))^\dagger,
    \end{align}
    where $\mathbf{y}_1 \defeq [y_1,\ldots,y_4]^T, \mathbf{y}_2 \defeq [y_5,\ldots,y_9]^T$.
    The helper nodes then send their quantum subsystems to the newcomer. Since each helper transmits one qudit, the total repair bandwidth is $d \beta_{\mathsf{q}} = 4$ qudits.
    
    Upon receiving the $4$ quantum subsystems, the newcomer performs the corresponding CSS stabilizer measurements. The resulting syndrome is
    \begin{align}
        \tilde{\mathbf{c}}_1 \defeq 
        \begin{bmatrix}
            \mathbf{s}_Z \\
            \mathbf{s}_X
        \end{bmatrix}
        =
        \begin{bmatrix}
            \mathbf{H}_X \mathbf{y}_2 \\
            \mathbf{H}_Z \mathbf{y}_1
        \end{bmatrix}.
    \end{align}
    Equivalently,
    \begin{align}
        \tilde{\mathbf{c}}_1 
        & =\begin{bmatrix}
            \mathbf{0}_{2\times 4} & \mathbf{H}_X \\
            \mathbf{H}_Z & \mathbf{0}_{2\times 4}
        \end{bmatrix} \mathbf{y} \\
        & = \begin{bmatrix}
            \mathbf{0}_{2 \times 2} & \mathbf{0}_{2 \times 2} & \mathbf{I}_{2} & \mathbf{I}_{2}\\
            \mathbf{I}_{2} & \mathbf{I}_{2} & \mathbf{0}_{2 \times 2} & \mathbf{0}_{2 \times 2}
        \end{bmatrix}
        \begin{bmatrix}
            (\tilde{\mathbf{\Lambda}} \tilde{\mathbf{V}})^{-1} & \mathbf{0}\\
            \mathbf{0} & (\tilde{\mathbf{\Lambda}}' \tilde{\mathbf{V}})^{-1}
        \end{bmatrix} \notag \\
        & \quad \times \begin{bmatrix}
            \tilde{\mathbf{\Lambda}} & \mathbf{0} \\
            \mathbf{0} & \tilde{\mathbf{\Lambda}}'
        \end{bmatrix}
        \begin{bmatrix}
            \tilde{\mathbf{V}} & \mathbf{0} \\
            \mathbf{0} & \tilde{\mathbf{V}}
        \end{bmatrix}
        \begin{bmatrix}
            \mathbf{M} \\
            \mathbf{M}'
        \end{bmatrix}
        \bar{\mathbf{v}}_1 \\
        & = \begin{bmatrix}
            \mathbf{S}_1' + \mathbf{S}_2' \\
            \mathbf{S}_1 + \mathbf{S}_2
        \end{bmatrix}
        \bar{\mathbf{v}}_1.
    \end{align}
    
    Finally, applying a permutation matrix to $\tilde{\mathbf{c}}_1$, we recover
    \begin{align}
        \begin{bmatrix}
            \mathbf{S}_1 + \mathbf{S}_2 \\
            \mathbf{S}_1' + \mathbf{S}_2'
        \end{bmatrix}
        \bar{\mathbf{v}}_1
        = \bar{\mathbf{c}}_1
    \end{align}
    which is exactly the content originally stored in the failed node $1$ in \eqref{eq:c_1}. Hence, node $1$ is exactly regenerated with $\alpha=d\beta_{\mathsf{q}} = 4$, achieving the optimal regenerating point.
\end{example}

\section{General Framework for Exact Regeneration}
In this section, we present a general achievability scheme for the optimal regenerating point under exact repair. We first provide a framework for $d = 2k-2$, and then extend it to $d > 2k-2$. Without loss of generality, we consider the normalized setting $\beta_{\mathsf q}=1$. Each node stores $\alpha = 2\alpha_0$ dits, partitioned into two instances of length $\alpha_0$. At the optimal regenerating point, we have $\alpha = d\beta_{\mathsf{q}} = B/k$.

\enlargethispage{-0.01in}
\subsection{Storage Design}
Consider the finite field $\mathbb{F}_q$ with $q \geq n$. Based on the product-matrix codes, we define the information matrix containing $B$ symbols as
\begin{align}
    \mathbf{M} = \begin{bmatrix}
        \mathbf{S}_1;
        \mathbf{S}_2
    \end{bmatrix}, \ \mathbf{M}' = \begin{bmatrix}
        \mathbf{S}_1';
        \mathbf{S}_2'
    \end{bmatrix} \in \mathbb{F}_{q}^{2 \alpha_0 \times \alpha_0},
\end{align}
where $\mathbf{S}_i$ and $\mathbf{S}_i'$, $i\in [2]$, are $\alpha_0 $ by $\alpha_0$ symmetric matrices, each containing $\alpha_0 (\alpha_0 + 1)/2$ distinct message symbols. The encoded data across the $n$ nodes is given by
\begin{align}
    \mathbf{C} = \mathrm{blkdiag}(\mathbf{V},\mathbf{V}) [\mathbf{M};\mathbf{M}'],
\end{align}
where $\mathbf{V} \in \mathbb{F}_q^{n \times 2 \alpha_0}$ is a Vandermonde matrix defined as
\begin{align}
    \mathbf{V} \defeq \begin{bmatrix}
        1& v_1& v_1^2 & \ldots& v_1^{2 \alpha_0-1}\\
        1& v_2& v_2^2 & \ldots& v_2^{2 \alpha_0-1}\\
        \vdots& \vdots&\vdots & \ddots& \vdots\\
        1& v_n&v_n^2 & \ldots& v_n^{2 \alpha_0-1}\\
    \end{bmatrix}.
\end{align}
Each row of $\mathbf{V}$, denoted as $\mathbf{v}_i^T$, can be decomposed as
\begin{align}
    \mathbf{v}_i^T = \begin{bmatrix}
         \Bar{\mathbf{v}}_i^T & \lambda_i \Bar{\mathbf{v}}_i^T
    \end{bmatrix},
\end{align}
where $\Bar{\mathbf{v}}_i \defeq [1 \ v_i \ \ldots \ v_i^{\alpha_0 - 1}]$ and $\lambda_i \defeq v_i^{\alpha_0}$. Node $i$ stores the $i$th and $(n+i)$th rows of $\mathbf{C}$, denoted as
\begin{align}
     \mathbf{v}_i^T \mathbf{M} & = \Bar{\mathbf{v}}_i^T \mathbf{S}_1 + \lambda_i \Bar{\mathbf{v}}_i^T \mathbf{S}_2, \\ \mathbf{v}_i^T \mathbf{M}' & = \Bar{\mathbf{v}}_i^T \mathbf{S}_1' + \lambda_i \Bar{\mathbf{v}}_i^T \mathbf{S}_2', 
\end{align}
respectively.

\subsection{Data-Retrieval}
The first instance $\mathbf{M}$ can be recovered from the storage of any $k$ nodes, i.e., $\mathbf{v}_{i_1}^T \mathbf{M}, \ldots, \mathbf{v}_{i_k}^T \mathbf{M}$, where $\{i_1,\ldots,i_k\} \subseteq [n]$, following the same scheme in \cite{Exact_regeneration}.
Similarly, $\mathbf{M}'$ can also be retrieved.

\subsection{Exact Regeneration}
Assume that node $f\in [n]$ fails, and the helper nodes are indexed by $s_{[d]} \subseteq [n]\setminus f$. The entanglement-assisted repair process proceeds as follows.
\begin{enumerate}
    \item \textbf{(Entanglement preparation)} The helper nodes share an entangled state $\ket{\psi} \in \mathrm{CSS}(\mathcal{C}_X,\mathcal{C}_Z)$, where the codes $\mathcal{C}_X$ and $\mathcal{C}_Z$ are defined via the parity-check matrices
    \begin{align}
        \mathbf{H}_X & \defeq \begin{bmatrix}
            \mathbf{I} & \lambda_f \mathbf{I}
        \end{bmatrix}
        (\tilde{\mathbf{\Lambda}}_2 \tilde{\mathbf{V}})^{-1},\label{eq:Hx}
        \\
        \mathbf{H}_Z & \defeq \begin{bmatrix}
            \mathbf{I} & \lambda_f \mathbf{I}
        \end{bmatrix}
        (\tilde{\mathbf{\Lambda}}_1 \tilde{\mathbf{V}} )^{-1},\label{eq:Hz}
    \end{align}
    respectively. Here, $\tilde{\mathbf{V}} \defeq \begin{bmatrix}
        \mathbf{v}_{s_1}^T;\ldots;\mathbf{v}_{s_d}^T
    \end{bmatrix}$ 
    consists of the storage matrix of the helper nodes and equivalently, $\tilde{\mathbf{V}} = \begin{bmatrix}
        \bar{\mathbf{V}} & \bar{\mathbf{\Lambda}}\bar{\mathbf{V}}
    \end{bmatrix}$, where $\bar{\mathbf{\Lambda}} \defeq \diag({\lambda}_{s_1},\ldots,{\lambda}_{s_d})$.
    Also,
    \begin{align}
        \tilde{\mathbf{\Lambda}}_1 & \defeq \diag(\mathbf{u}) (-\lambda_f \mathbf{I} + \bar{\mathbf{\Lambda}})^{-1}, \\
        \tilde{\mathbf{\Lambda}}_2 & \defeq \diag(\mathbf{u}')(-\lambda_f \mathbf{I} + \bar{\mathbf{\Lambda}})^{-1},
    \end{align}
    $\mathbf{u} = [u_1,\ldots,u_d]$ and $\mathbf{u}' = [u_1',\ldots,u_d']$ are designed as $u_i \neq 0, \ \forall i \in [d]$,
    \begin{align}
        u_j' \defeq \frac{1}{u_j}
        \left(\prod_{i \in [d],\, i \neq j}(u_j - u_i)\right)^{-1}, \  \forall j \in [d].\label{eq:GRS}
    \end{align}
    To ensure a valid CSS construction, we need the dual-containment condition $\mathcal{C}_X^{\perp} \subseteq \mathcal{C}_Z$, which is equivalent to $\mathbf{H}_X \mathbf{H}_Z^T = \mathbf{0}$ \cite{calderbank1996good,steane1996multiple}. This is verified by the following lemma.
    
    \begin{lemma}
        The parity-check matrices $\mathbf{H}_X$ and $\mathbf{H}_Z$ defined in \eqref{eq:Hx} and \eqref{eq:Hz} satisfy
        \begin{align}
            \mathbf{H}_X \mathbf{H}_Z^T = \mathbf{0}.
        \end{align}
    \end{lemma}
    
    \begin{Proof}
    Note that, we can rewrite $\mathbf{H}_Z$ as
    \begin{align}
        \mathbf{H}_Z &= \begin{bmatrix}
            \mathbf{I} & \lambda_f \mathbf{I}
        \end{bmatrix}
        (\tilde{\mathbf{\Lambda}}_1 \tilde{\mathbf{V}})^{-1}\\
        & = \begin{bmatrix}
            \mathbf{0} & \mathbf{I}
        \end{bmatrix}
        \begin{bmatrix}
            \mathbf{0} & \mathbf{I} \\
            \mathbf{I} & \lambda_f \mathbf{I} 
        \end{bmatrix}
        (\tilde{\mathbf{\Lambda}}_1 \tilde{\mathbf{V}})^{-1} \\
        & = \begin{bmatrix}
            \mathbf{0} & \mathbf{I}
        \end{bmatrix}  
        \begin{bmatrix}
            -\lambda_f \mathbf{I} & \mathbf{I} \\
            \mathbf{I} & \mathbf{0}
        \end{bmatrix}
        ^{-1}(\tilde{\mathbf{\Lambda}}_1 \tilde{\mathbf{V}})^{-1}\\
        & = \begin{bmatrix}
            \mathbf{0} & \mathbf{I}
        \end{bmatrix} \left( \tilde{\mathbf{\Lambda}}_1 \tilde{\mathbf{V}} 
        \begin{bmatrix}
            -\lambda_f \mathbf{I} & \mathbf{I} \\
            \mathbf{I} & \mathbf{0}
        \end{bmatrix}
        \right)^{-1} \\
        & = \begin{bmatrix}
            \mathbf{0} & \mathbf{I}
        \end{bmatrix}
        \begin{bmatrix}
            (-\lambda_f \mathbf{I} + \bar{\mathbf{\Lambda}} )\tilde{\mathbf{\Lambda}}_1 \bar{\mathbf{V}} & \tilde{\mathbf{\Lambda}}_1 \bar{\mathbf{V}}
        \end{bmatrix}^{-1}, 
    \end{align}
    where the last equality is because $\tilde{\mathbf{V}} = \begin{bmatrix}
        \bar{\mathbf{V}} & \bar{\mathbf{\Lambda}}\bar{\mathbf{V}}
    \end{bmatrix}$.
    
    Then, we have
    \begin{align}
        \mathbf{H}_Z \mathbf{H}_X^T 
        & = \begin{bmatrix}
            \mathbf{0} & \mathbf{I}
        \end{bmatrix} \mathbf{A}^{-1}
        \begin{bmatrix}
            \mathbf{0} \\
            \mathbf{I}
        \end{bmatrix},\label{eq:HZX}
    \end{align}
    where 
    \begin{align}
         \mathbf{A}&\defeq  
        \begin{bmatrix}
            \bar{\mathbf{V}}^T \tilde{\mathbf{\Lambda}}_2 (-\lambda_f \mathbf{I} + \bar{\mathbf{\Lambda}}) \\
            \bar{\mathbf{V}}^T \tilde{\mathbf{\Lambda}}_2
        \end{bmatrix}
        \begin{bmatrix}
            (-\lambda_f \mathbf{I} + \bar{\mathbf{\Lambda}} )\tilde{\mathbf{\Lambda}}_1 \bar{\mathbf{V}} & \tilde{\mathbf{\Lambda}}_1 \bar{\mathbf{V}}
        \end{bmatrix}\\
        & = 
        \begin{bmatrix}
            \bar{\mathbf{V}}^T \diag(\mathbf{u}') \\
            \bar{\mathbf{V}}^T \tilde{\mathbf{\Lambda}}_2
        \end{bmatrix}
        \begin{bmatrix}
            \diag(\mathbf{u})\bar{\mathbf{V}} &
            \tilde{\mathbf{\Lambda}}_1 \bar{\mathbf{V}}
        \end{bmatrix} \\
        & = 
        \begin{bmatrix}
            \mathbf{0} & \bar{\mathbf{V}}^T \diag(\mathbf{u}') \tilde{\mathbf{\Lambda}}_1 \bar{\mathbf{V}}\\
            \bar{\mathbf{V}}^T \tilde{\mathbf{\Lambda}}_2 \diag(\mathbf{u}) \bar{\mathbf{V}} & \bar{\mathbf{V}}^T \tilde{\mathbf{\Lambda}}_2 \tilde{\mathbf{\Lambda}}_1 \bar{\mathbf{V}}
        \end{bmatrix},\label{eq:A}
    \end{align}
    where the top-left block in \eqref{eq:A} is $\mathbf{0}$ because $\bar{\mathbf{V}}^T \diag(\mathbf{u}') \diag(\mathbf{u}) \bar{\mathbf{V}} = \mathbf{0}$ by the design in \eqref{eq:GRS}. Observe that all nonzero blocks in \eqref{eq:A} are invertible. Thus, the bottom-right block of $\mathbf{A}^{-1}$ is $\mathbf{0}$. Hence, the $\mathbf{H}_Z \mathbf{H}_X^T $ in \eqref{eq:HZX} is $ \mathbf{0}$, which implies $\mathbf{H}_X \mathbf{H}_Z^T = \mathbf{0}$.
    \end{Proof}

    \item \textbf{(Encoding at helper nodes)}
    Each helper node $s \in s_{[d]}$ first computes
    \begin{align}
        \lambda_s \mathbf{v}_{s}^T [\mathbf{S}_1;\mathbf{S}_2] \Bar{\mathbf{v}}_f, \quad \lambda_s'\mathbf{v}_{s}^T [\mathbf{S}_1';\mathbf{S}_2'] \Bar{\mathbf{v}}_f.
    \end{align}
    Collecting the symbols from all helper nodes, the classical information to be encoded into the quantum system is 
    \begin{align}
        \mathbf{y} = \begin{bmatrix}
            \mathbf{y}_1 \\
            \mathbf{y}_2
        \end{bmatrix} \defeq 
        \begin{bmatrix}
            \tilde{\mathbf{\Lambda}}_1 & \mathbf{0} \\
            \mathbf{0} & \tilde{\mathbf{\Lambda}}_2
        \end{bmatrix}
        \underbrace{\begin{bmatrix}
            \tilde{\mathbf{V}} & \mathbf{0} \\
            \mathbf{0} & \tilde{\mathbf{V}}
        \end{bmatrix}
        \begin{bmatrix}
            \mathbf{M}\\
            \mathbf{M}'
        \end{bmatrix}}_{\mathrm{storage}\ \mathrm{at}\ \mathrm{helper}\ \mathrm{nodes}}\bar{\mathbf{v}}_f,
    \end{align}
    where $\tilde{\mathbf{\Lambda}}_1 = \diag(\tilde{\lambda}_1,\ldots,\tilde{\lambda}_{\alpha_0})$, $\tilde{\mathbf{\Lambda}}_2 = \diag(\tilde{\lambda}_1',\ldots,\tilde{\lambda}_{\alpha_0}')$, and
    $\mathbf{y}_i \in \mathbb{F}_q^{\alpha_0 \times 1}, i\in [2],$ is the $i$th instance.
    Then, each helper node applies $\mathsf{X}(y_i) \mathsf{Z}(y_{i+d})$ to its local quantum system, resulting in the quantum system becoming
    \begin{align}
        \rho^{\rm enc} = \mathsf{X}(\mathbf{y}_1)\mathsf{Z}(\mathbf{y}_2) \ket{\psi}\bra{\psi} (\mathsf{X}(\mathbf{y}_1)\mathsf{Z}(\mathbf{y}_2))^\dagger.
    \end{align}

    \item \textbf{(Exact generation)}
    Upon receiving the quantum systems, the newcomer performs the stabilizer measurement. The obtained syndrome is
    \begin{align}
        \begin{bmatrix}
            \mathbf{0} & \mathbf{H}_X \\
            \mathbf{H}_Z & \mathbf{0}
        \end{bmatrix} \mathbf{y} & =
        \begin{bmatrix}
             \mathbf{0} & \mathbf{0} & \mathbf{I} & \lambda_f \mathbf{I} \\
             \mathbf{I} & \lambda_f \mathbf{I} &\mathbf{0} &\mathbf{0}
        \end{bmatrix}
        \begin{bmatrix}
            \mathbf{M}\\
            \mathbf{M}'
        \end{bmatrix}\bar{\mathbf{v}}_f\\
        & = \begin{bmatrix}
            \mathbf{S}_1' \bar{\mathbf{v}}_f + \lambda_f \mathbf{S}_2' \bar{\mathbf{v}}_f \\
            \mathbf{S}_1 \bar{\mathbf{v}}_f + \lambda_f \mathbf{S}_2 \bar{\mathbf{v}}_f
        \end{bmatrix}.
    \end{align}
    By applying a permutation that swaps these two blocks, the newcomer recovers the content  stored in the failed node $f$.
\end{enumerate}

\subsection{Extension to $d > 2k-2$}
For $d > 2k-2$, we can still achieve the optimal regenerating point by increasing $\beta_{\mathsf{q}}$ appropriately. The key observation is that the repair procedure above only requires a set of $2k-2$ helpers. With $d$ available helpers, we can partition the original file into $\binom{d}{2k-2}$ independent sub-files, each encoded using the above scheme with a distinct subset of $2k-2$ helpers. Each helper then participates in several such sub-schemes and transmits one qudit per sub-scheme, resulting in $\beta_{\mathsf{q}} = \binom{d}{2k-2}$. The total storage per node becomes $\alpha = \binom{d}{2k-2} \bar{\alpha}$ with $\bar{\alpha} = 2k-2$ being the storage for each subfile, and the total file size is $B = \binom{d}{2k-2} \bar{B}$, where $\bar{B} = k \bar{\alpha}$ is the size of each sub‑file.

\bibliographystyle{unsrt}
\bibliography{Ref1.bib}

\end{document}